\documentclass[aps,prl,floats,floatfix,twocolumn]{revtex4-1}

\usepackage{color} 
\usepackage{latexsym}
\usepackage{amsmath} 
\usepackage{amssymb} 
\usepackage{bm}
\usepackage{wasysym}
\usepackage[normalem]{ulem}
\usepackage{mathtools}

\usepackage{graphicx}
\usepackage{epstopdf}
\graphicspath{{./}{./Figs/}}
\usepackage[percent]{overpic}

\usepackage[colorlinks,linkcolor=blue,urlcolor=blue,bookmarks=true,pdftitle={}]{hyperref}
\newcommand{\hide}[1]{}

\newcommand{\Fig}[1] {\ref{#1}}

\makeatother

\begin{document}

\title{Chiral states in coupled-lasers lattice by on-site complex potential}

\author{Sagie Gadasi$^{\dagger}$, Geva Arwas$^{\dagger}$, Igor Gershenzon, Asher Friesem and Nir Davidson}

\affiliation{
Department of Physics of Complex Systems, Weizmann Institute of Science, Rehovot 7610001, Israel
}

\begin{abstract}

The ability to control the chirality of physical devices is of great scientific and technological importance, from investigations of topologically protected edge states in condensed matter systems to wavefront engineering, isolation, and unidirectional communication.
When dealing with large networks of oscillators, the control over the chirality of the bulk states becomes significantly more complicated and requires complex apparatus for generating asymmetric coupling or artificial gauge fields. 
Here we present a new approach for precise control over the chirality of a triangular array of hundreds of symmetrically-coupled lasers, by introducing a weak non-Hermitian complex potential, requiring only local on-site control of loss and frequency.
In the unperturbed network, lasing states with opposite chirality (staggered vortex and staggered anti-vortex) are equally probable. We show that by tuning the complex potential to an exceptional point, a nearly pure chiral lasing state is achieved.
While our approach is applicable to any oscillators network, we demonstrate how the inherent non-linearity of the lasers effectively pulls the network to the exceptional point, making the chirality extremely resilient against noises and imperfections. 
\end{abstract}

\maketitle

\section{Introduction}
Chirality is a fundamental property of nature, and the ability to control it is of great scientific and technological importance. In physical systems chiral modes are typically induced by a magnetic field or by its analogs. For example, in quantum Hall effect, chiral edge currents are generated by magnetic fields  \cite{QHE}, in ultra-cold atoms chiral edge states are induced by artificial gauge fields \cite{Mancini1510,Lin2009,Dalibard2011}, and the rotation direction of hurricanes, depends on the their location on earth, as a result of Coriolis force. 
The underlying mechanism in all of these examples, is the breaking of time reversal symmetry of the system, which results in an energy splitting of two, otherwise degenerate, oppositely propagating modes.

In optics, the Faraday effect, induced by a magnetic field, is widely used for breaking time reversal symmetry and to induce directionality \cite{FE1}. However, the required arrangement is bulky and miniaturizing it to chip-scale size is difficult.
Nevertheless, thanks to gain saturation, unidirectional operation of miniature ring lasers can be achieved by asymmetric coupling between the two counter-propagating modes  \cite{Craft1993,Peng6845,Miao464,Zhang760}.
When dealing with large network of optical oscillators, the control over the chirality of the supermode is done by asymmetric coupling between the individual oscillators. For instance, synthetic magnetic fields allowed  the observation of chiral edge states in coupled waveguides \cite{Hafezi2013} and the generation of topological insulator lasers \cite{TIL}. 
While the principle of operation of these methods is intuitive and well-studied, asymmetric coupling is usually interferometrically sensitive, requires complicated fabrication facilities, and its implementation in different physical systems is not straight forward.

In this paper we present a new approach for controlling the chirality of the supermode of an arbitrarily large network of nonlinear oscillators by resorting to a non-Hermitian on-site complex potential.
Herein, complex potential refers to the combination of real (frequency detuning) and imaginary (loss) terms on the diagonal of the Hamiltonian that describes the system.
The complex potential generates an asymmetric energy flow between supermodes with opposite chirality, making one more favorable than the other. Remarkably, the asymmetric energy flow results only from on-site potentials (diagonal) while the coupling between the oscillators  (off-diagonal) remains strictly symmetric, without generating synthetic magnetic fields as in \cite{TIL}.   

While this approach can be applied to any network of nonlinear oscillators, we demonstrate it with a triangular lattice of hundreds of symmetrically-coupled lasers, in which chiral symmetry is manifested by a two-fold loss-degeneracy of its fundamental lasing mode.
The loss-degeneracy implies that the two supermodes are equally probable upon lasing, and therefore the chirality of the system is zero on average. 
We show that by tuning the complex potential to an exceptional point (EP) of the effective Hamiltonian, it is possible to reach a lasing mode with nearly pure chirality. 
We also show that the nonlinearity of the lasers pulls the system towards the EP, such that the chiral states are extremely robust to noises in disorder.

The presented approach is scalable and can be applied to any number of lasers, to a variety of network geometries and to other coupled lasers lattices, such as diode-laser and vertical-cavity-surface-emitting-laser (VCSEL) arrays. Requiring only local on-site control of loss and frequency, the method can be readily applied to force chirality on any physical systems of coupled oscillators, such as microwave resonators \cite{Sliwa2015}, cold atoms lattices \cite{Gross995}, mechanical systems \cite{Ssstrunk47}, and even human networks \cite{Shahal2020}.

\begin{figure*}[!ht]

\includegraphics[width=1.0\hsize]{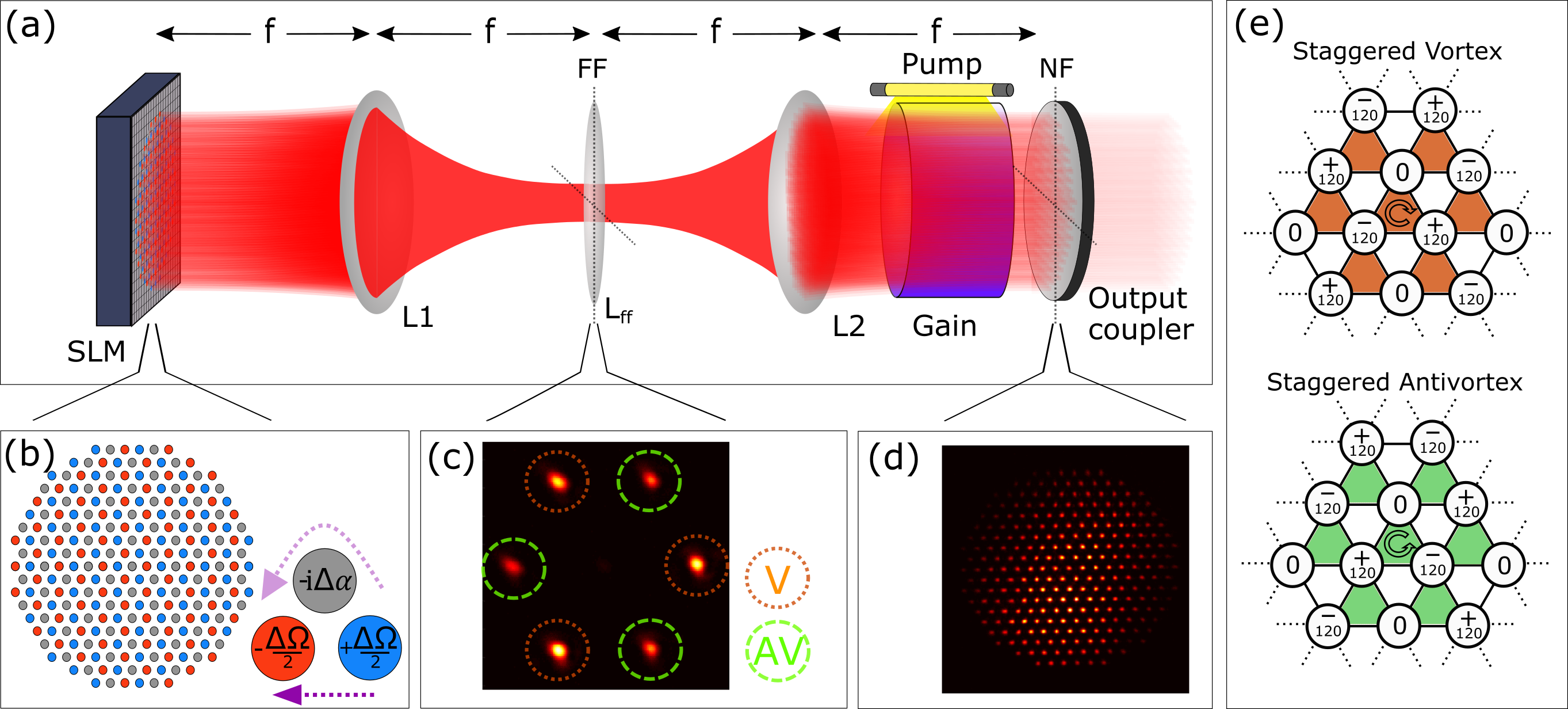}

\caption{ \label{fg1} \textbf{Experimental arrangement and lasing modes.} (a) Schematic arrangement of the digital degenerate cavity laser. (b) The lasers network geometry and  complex potential that were applied by the SLM. In each triangle, loss of $\Delta\alpha$ was applied to one of the lasers, and a detuning of $\pm\frac{\Delta\Omega}{2}$ was applied to the other two. (c) A measured far-field image of the triangular laser lattice. The spots mark by V (AV) correspond to staggered vortex (antivortex) lasing mode. The presence of both the V and AV spots is due to the existence of multiple longitudinal modes. (d) A detected near-field image of the lasers indicating local uniformity of their intensity.  (e) Illustration of the two loss-degenerate lasing modes of a triangular lattice of negatively-coupled lasers, the staggered vortex and staggered antivortex.}
\end{figure*}

\section{Experimental arrangement and results}

The approach can be heuristically explained. 
When two coupled oscillators have a small frequency detuning, they oscillate with a phase difference that is dictated by the detuning \cite{Kuramoto}.
The frequency detuning induces directionality on the oscillators' phases.
For a closed a ring of oscillators, one might attempt to achieve chirality by introducing a frequency detuning between two oscillators to “enforce” a directional phase gradient.
However, for uniform oscillation amplitudes, the detuning actually induces opposite phase gradients on the direct link and on the indirect link, and therefore the directionality is canceled out and chiral symmetry is maintained \cite{Supp}.
We thus propose to recover the directionality and break the chiral symmtery by weakening the indirect link, by introducing loss to one of the other oscillators (inset of figure \Fig{fg1}(b)). We show that an introduction of frequency detuning and loss, a complex potential, indeed induces  chirality to an oscillators network (\cite{detunChirality}), and pure chirality is achieved when the complex potential is tuned to an exception point.

To demonstrate control over the chirality of an oscillators network, we use a digital degenerate-cavity laser (DDCL) \cite{Tradonskyeaax4530} to set up a triangular lattice of 253 negatively-coupled lasers \cite{Supp}. The DDCL,  schematically shown in figure \Fig{fg1}(a), is comprised of a 4f telescope, an Nd:YAG gain medium, a coupling element, and a spatial light modulator (SLM). The gain medium is pumped by a $200\mu s$ pulsed Xenon flash lamp with $1Hz$ repetition rate. The SLM serves as a digital mirror, enabling the generation of laser networks with arbitrary geometry, and accurately controlling the loss and frequency detuning (phase) of each of the lasers separately (as in figure \Fig{fg1}(b)). Negative and symmetric coupling of $\kappa=75$MHz between all nearest-neighbor lasers is achieved by placing a lens in the far-field (FF) plane inside the cavity \cite{Supp}. The FF lens alters the perfect imaging of the 8f telescope and scatters light from each laser to its neighbors \cite{Mahler2019}. 
An external imaging configuration images the near-field (NF) plane and its Fourier transform, the FF plane. In each laser pulse of the DDCL, many uncoupled longitudinal modes lase simultaneously \cite{Simon2020}. Hence, effectively, the detected NF and FF distributions in each lasing pulse are ensemble averages over many independent realizations of the experiment.

Typical FF and NF intensity distributions of the triangular lattice of identical lasers are presented in figures \Fig{fg1}(c) and \Fig{fg1}(d). The two degenerate lasing supermodes are the staggered vortex (SV) and the staggered anti-vortex (SAV) modes, illustrated in figure \Fig{fg1}(e). The chirality of a lasing mode is encoded in the FF intensity distribution, provided that the lasers intensities are locally uniform (figure 1(d)). The FF of the SV (SAV) mode contains the three spots marked by V (AV) in figure  \Fig{fg1}(c). We quantify the chirality $c$ of the lasing modes by: 
\begin{equation}\label{chirality_eqn}
c=\frac{I_{V}-I_{AV}}{I_{V}+I_{AV}}\,,
\end{equation}
where $I_{V}$ and $I_{AV}$ are the average intensities of the V and AV spots. 

\begin{figure}

\includegraphics[width=0.98\hsize]{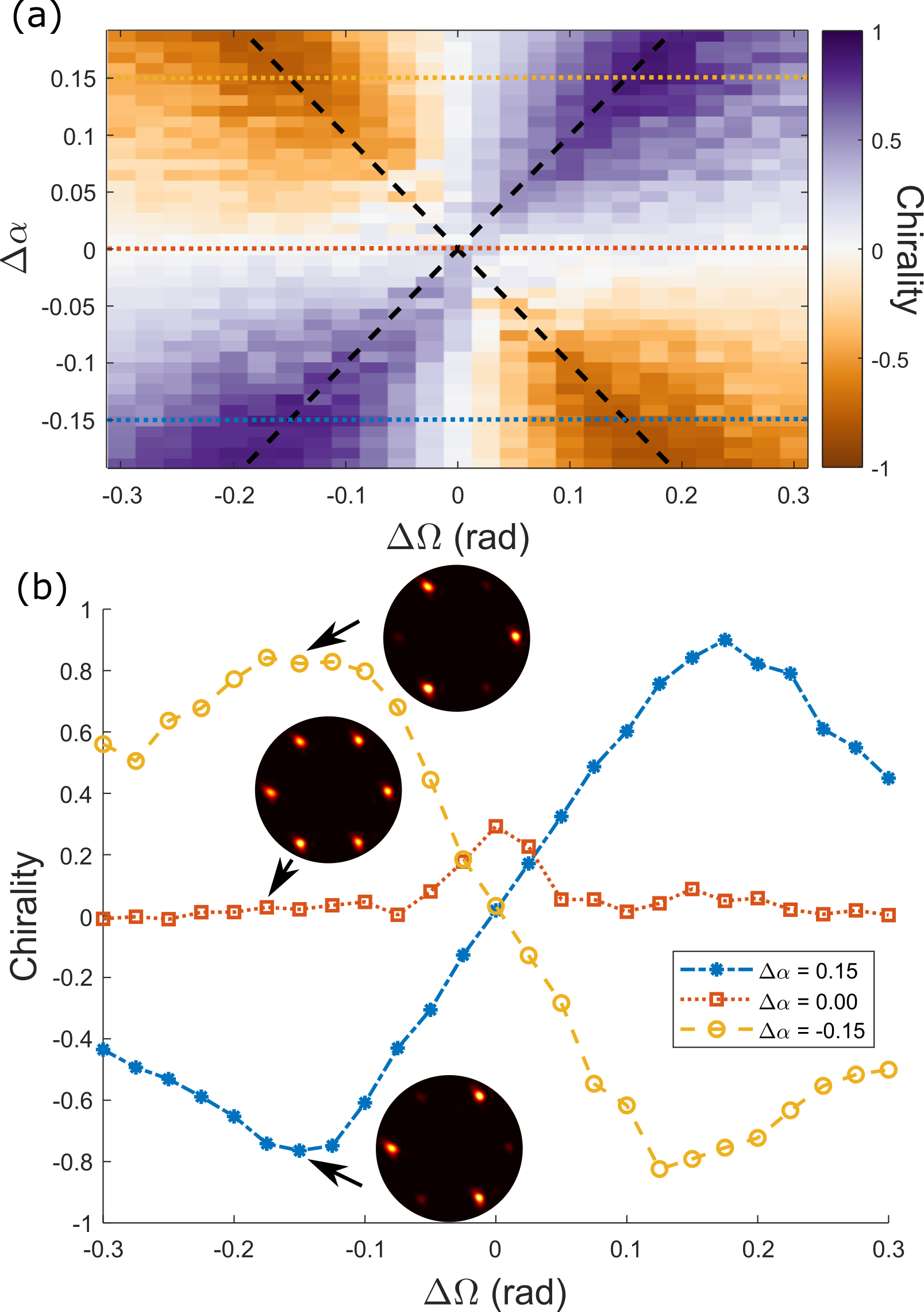}
\caption{ \label{fg2} \textbf{Experimentally measured chirality, induced by complex potential.}  (a) The measured chirality $c$ of the lasing mode as a function of frequency detuning $\Delta\Omega$ and relative loss $\Delta\alpha$. Maximal chirality is obtained along the lines $\Delta\Omega=\Delta\alpha$, marked by the black dashed lines. (b) Cross sections of the chirality for varying $\Delta\Omega$ at fixed $\Delta\alpha$ values (marked by the dotted colored lines in (a)). The measurement points are connected by lines. Insets show typical far-field images at selected points.}
\end{figure}

The loss and frequency detuning of the lasers in the array were modulated according to figure \Fig{fg1}(b) to form the complex potential. For each triangle in the lattice we added a loss of $\Delta\alpha$ to one of the lasers, and applied a frequency detuning of $\pm\frac{1}{2}\Delta\Omega$ to the other two lasers \cite{detuning}. The loss and detuning were applied through the complex reflectivity of the SLM for each laser $R=e^{-\Delta\alpha+i\Delta\Omega}$. 
The values of the detuning $\Delta\Omega$ and loss $\Delta\alpha$ were varied and the chirality, calculated from the measured FF, is presented in figure  \Fig{fg2}(a). As evident, applying frequency detuning alone ($\Delta\alpha=0$) or loss alone ($\Delta\Omega=0$) is not sufficient to induce chirality, in agreement with our heuristic explanation above. We also find that maximal chirality is obtained when the detuning and loss are about equal (marked by the black dashed lines).
Three loss cross sections from the two-dimensional diagram are presented in figure \Fig{fg2}(b). When loss of $\Delta\alpha=0.15$ is applied, small detuning is already enough to break the chiral symmetry of the system. Increasing the detuning results in the increase in chirality up to the point $\Delta\alpha\approx\Delta\Omega$, where nearly pure chirality is obtained. Increasing the detuning further leads to a decrease in the system's chirality. If instead of loss, gain of $\Delta\alpha=-0.15$ is applied, we observe a similar behavior but with the opposite chirality for the same detuning. When no loss is applied ($\Delta\alpha=0$) and only the detuning is varied, there is no significant change in the system's chirality. Selected FF intensity distributions are presented as insets in figure \Fig{fg2}(b). The sharp Bragg peaks in the FF measurements indicate that the phase locking of the coupled lasers is nearly perfect throughout the array.

\section{Theory and discussion}

\begin{figure}
\includegraphics[width=0.98\hsize]{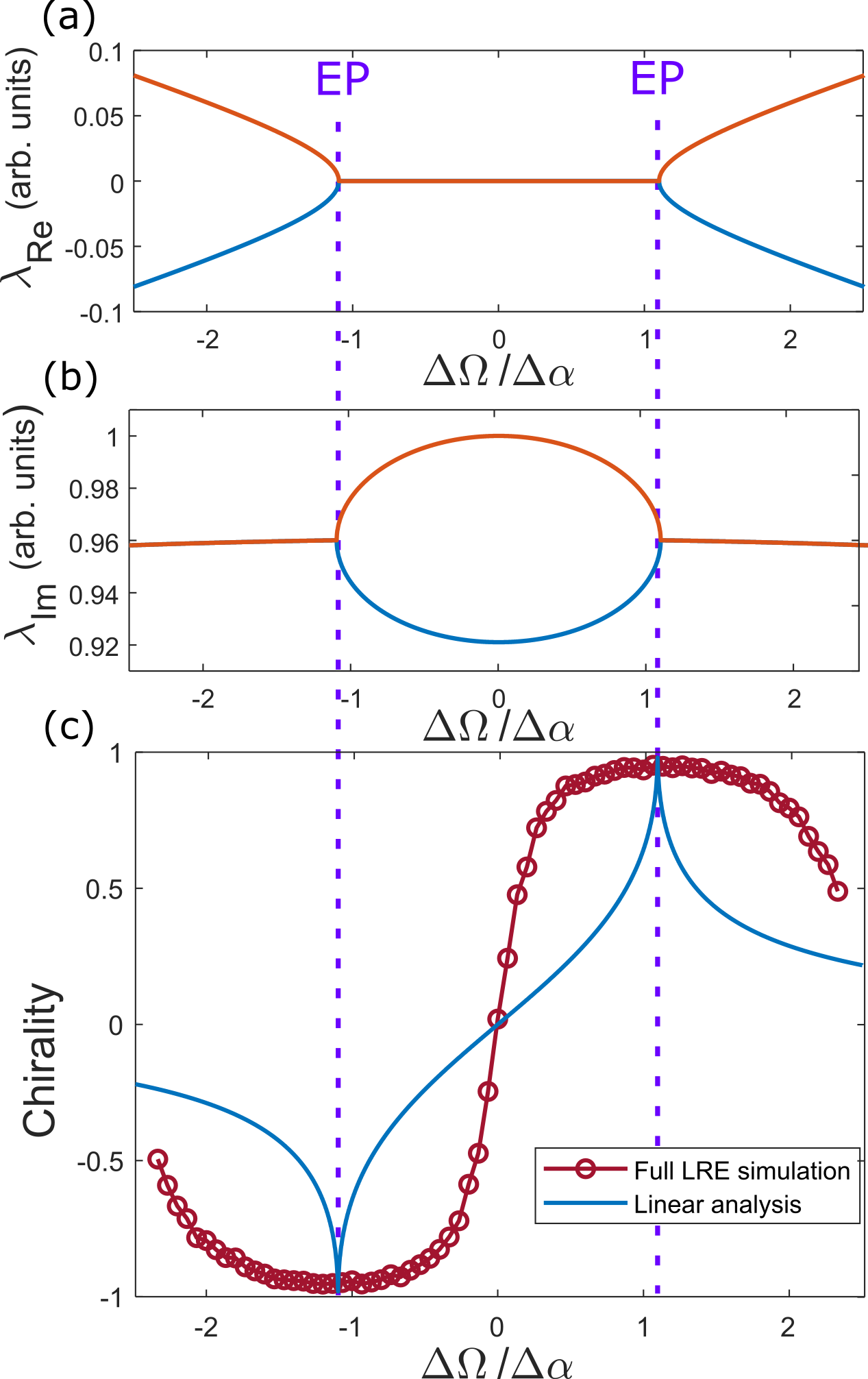}
\caption{ \textbf{Theoretical analysis.}  \label{fg3} (a) and (b) display the real and imaginary parts of the eigenvalues of the systems' minimal loss cold-cavity modes as a function of $\Delta\Omega$ and fixed $\Delta\alpha$. Exceptional points, marked by the vertical purple dashed lines, emerge at $\Delta\Omega_{\text{EP}}\approx1.1\Delta\alpha$ (for $\Delta\alpha=0.12\kappa$), where square-root splitting is apparent. (c) The solid blue curve presents the chirality of the minimal-loss cold-cavity mode. The red circles display the chirality of the laser network, obtained from simulation of the full nonlinear laser rate equations. The data points are connected by a solid line. }
\end{figure}

To explain our experimental results, we begin by analyzing the cold-cavity modes of the linear part of the system (i.e. the modes of cavity without the nonlinear gain). 
The complex potential breaks the lattice into three sublattices, each by itself has translation symmetry. 
Hence, by neglecting finite-size corrections the lattice can be described by an analogous system made of one unit cell of three cavities with a modified coupling coefficient of $\kappa = 3 \tilde{\kappa}$, where $\tilde{\kappa}$  is the coupling strength between two neighboring lasers in the actual lattice.
In this linear regime, the evolution of the electric field in the cavities of one unit cell is governed by a Schrodinger-like equation,
 $\frac{d\boldsymbol{E}}{dt}=-i\mathcal{H}\boldsymbol{E}$, where 
\begin{equation}
\mathcal H=\left(\begin{array}{ccc}
-\frac{1}{2}\Delta\Omega & -i\kappa & -i\kappa\\
-i\kappa & -i\Delta\alpha & -i\kappa\\
-i\kappa & -i\kappa & +\frac{1}{2}\Delta\Omega
\end{array}\right)\,\label{eq:three_lasers_hamiltonian}
\end{equation}
is the Bloch Hamiltonian, and $\boldsymbol{E}$ is a column vector of the electric field in the three cavities \cite{Supp}. The imaginary coupling coefficients correspond to a dissipative (non energy conservative) coupling \cite{linearSteadyState,PhysRevApplied.12.054039}. 
Without $\Delta\Omega$ and $\Delta\alpha$, the eigenmodes of the system are the vortex, antivortex and uniform phase mode (see \cite{Supp}). 
The introduction of $\Delta\Omega$ or $\Delta\alpha$ breaks the invariance of the system for rotation in $\frac{2\pi}{3}$, and therefore the vortex and antivortex modes are no longer eigenmodes of the system.
The loss and frequency associated with each of the system's new eigenmodes are given by the imaginary and real parts of the corresponding eigenvalues $\{\lambda\}$, and are presented in figures \Fig{fg3}(a) and \Fig{fg3}(b) for fixed $\Delta\alpha$, and varying $\Delta\Omega$. The transition of the eigenvalues from being purely imaginary to being purely real is attributed to the system being anti-PT symmetric \cite{Supp,Peng2016}. 
An EP \cite{Mirieaar7709}, a non-Hermitian degeneracy at which the two eigenmodes become identical, emerges when the detuning is adjusted to specific values: $\Delta\Omega_{\text{EP}}\approx\Delta\alpha$ \cite{Supp}. At the EP, the two eigenmodes collapse into a nearly pure vortex or antivortex mode\cite{Supp}.
The chirality of the eigenmodes, calculated from their projection on the vortex and antivortex modes and according to equation \ref{chirality_eqn}, is displayed in figure \Fig{fg3}(c). When a relative loss of $\Delta\alpha$ is applied, and the relative frequency detuning $\Delta\Omega$ is raised from zero, chirality emerges and increases up to the EP, where the eigenmodes become pure vortex or antivortex. Increasing $\Delta\Omega$ further, results in a decrease in the chirality of the eigenmodes.

\begin{figure}
\includegraphics[width=0.98\hsize]{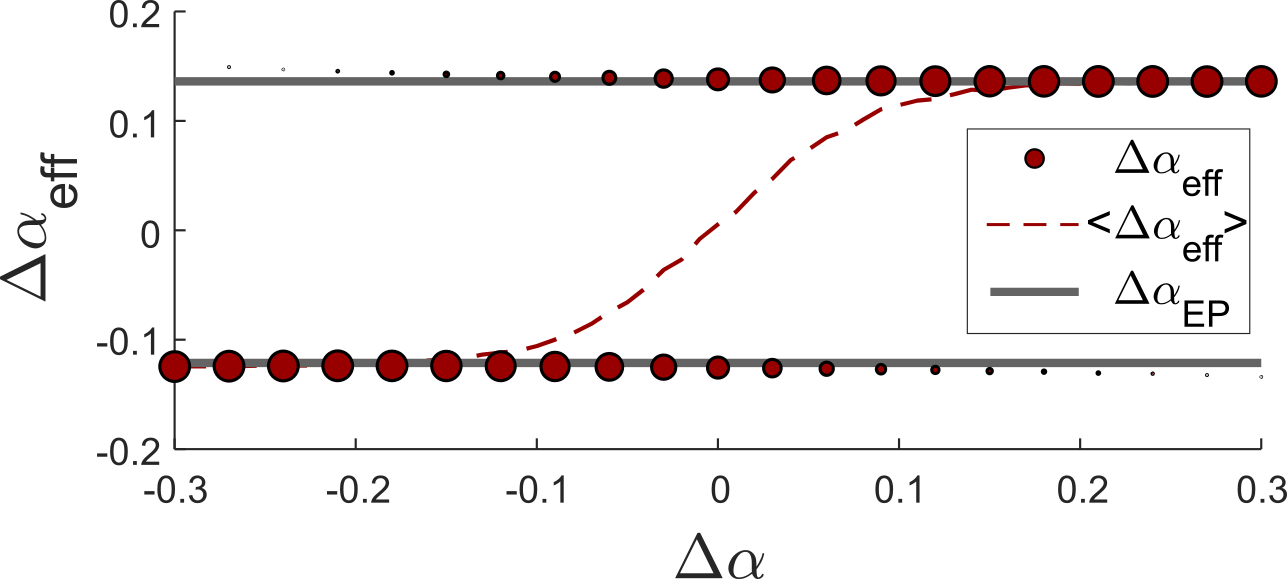}
\caption{
Simulation results for the effective loss $\Delta\alpha_{\text{eff}}$ (red circles) as a function of the applied loss $\Delta\alpha$ for $\Delta\Omega=0.15$rad and unsaturated gain of $g_0=1.2g_{0,\text{th}}$, where $g_{0,\text{th}}$ is the unsaturated gain at the lasing threshold.  The circles' diameter represents the prevalence of $\Delta\alpha_{\text{eff}}$ among the 5000 realizations. The dashed line shows an ensemble average $\langle\Delta\alpha_{\text{eff}}\rangle$. 
It is clearly seen that the steady-state $\Delta\alpha_{\text{eff}}$ approaches the values of $\Delta\alpha_{\text{EP}}$ (horizontal lines). This demonstrates the effect that the nolinearity of the lasers pulls the system to the EPs. Similar results  obtained for other values of $\Delta\Omega$ and $g_0$  are presented in the supplementary information.
\label{fg4}}
\end{figure}

Juxtaposing the cold-cavity linear-analysis results and the experimental results reveals a very good correspondence. The linear-analysis explains the emergence of chirality, and its behavior as a function of the applied loss and detuning. Also, evidence of the EPs at $\Delta\Omega_{\text{EP}}\approx\pm\Delta\alpha$, where the eigenmodes have nearly pure chirality are present in figure \Fig{fg2}(a) at the predicted locations. 
Nevertheless, the experimental chirality peaks in figure \Fig{fg2}(b) are significantly broadened relative to the linear theory (figure \Fig{fg3}(c)). This is also manifested in the broad chiral lines along $\Delta\Omega=\pm\Delta\alpha$ in figure \Fig{fg2}(a), where small changes in the detuning or loss have small effect on the chirality of the lasing mode.  We show below that  this striking difference emerges from the nonlinearity of the system. 

A more complete description of the system is obtained by numerically simulating the full nonlinear laser rate equations (LRE) \cite{PhysRevLett.92.093905}. 
Simulations results reveal that the lasing mode at steady-state is either vortex or antivortex modes, rather than a superposition of the two, and that the outcome of varying the value of $\Delta\Omega$ for a fixed $\Delta\alpha$ is a change in the probability to get one mode over the other \cite{Supp}. 
The red curve in figure \Fig{fg3}(c) show the chirality obtained from numerical simulations of the laser rate equations, averaged over 5000 realizations. The simulations reproduces the broad pure-chirality peaks around the EPs, similarly to the experimental results.

To explain intuitively the mechanism of this surprising effect, which is a consequence of the nonlinear gain, we now analyze how it modifies the effective loss between the lasers and enables convergence to either vortex or antivortex modes irrespective of the applied $\Delta\alpha$ \cite{Supp}.
At steady-state, the nonlinear gain of the $i$'th laser is given by ${G_i =  \frac{g_{0,i}}{1+\frac{|E_i|^2}{I_{\text{sat}}}}}$, where $g_{0,i}$ is the laser's unsaturated gain, dictated by the pumping rate and $I_{\text{sat}}$ is the saturation intensity of the gain medium. 
When the pumping rates are uniform for all the lasers, the nonlinear gain is larger for lasers with lower intensities, and therefore, at steady state, the effective loss between the lasers is given by:
\begin{equation}
\Delta\alpha_{\text{eff}} = \Delta\alpha - \Delta G  \, ,
\label{eq:ss_gain}
\end{equation}
where $\Delta G$ is the gain difference between the lossy laser and the other two lasers.
The nonlinear gain therefore pulls the system to steady-state intensities that are more homogeneous than those of the corresponding linear system.
Remarkably, the simulation results reveal that the effective loss $\Delta\alpha_{\text{eff}}$ is exactly the loss that is required to bring the system to an EP for the applied detuning ($\Delta\alpha_{\text{EP}}$). 
The effect is demonstrated in figure \ref{fg4}, where $\Delta\alpha_{\text{eff}}$ is displayed as a function of the applied $\Delta\alpha$ for $\Delta\Omega=0.15$rad.
The red circles display $\Delta\alpha_{\text{eff}}$, and their diameter represent their prevalence among the 5000 realizations. 
The prevalence is also shown by the ensemble average $\langle\Delta\alpha_{\text{eff}}\rangle$ (dashed line).
It is clearly seen that the steady-state $\Delta\alpha_{\text{eff}}$ approaches the value of $\Delta\alpha_{\text{EP}}$  (dotted line). 
Simulation results for other values of $\Delta\Omega$ and pumping rates  \cite{Supp} confirm that in general the nonlinearity of the lasers pulls the system towards the EP \cite{Nonlinear1,Nonlinear2}. 
The nonlinearity ensures that inducing chirality by complex potential is extremely robust against noises and perturbations, making this approach appealing for practical applications and large scale fabrication.

\section{Conclusions}

In this work we showed experimentally that the introduction of an on-site complex potential to a triangular lattice of hundreds of symmetrically-coupled lasers, drives the system to a well-controlled pure chiral supermode of staggered vortex or anti vortex modes. We also showed that the chirality is maximized when the complex potential is tuned to specific values, which correspond to the EPs of the linear system. At the EPs the eigenmodes of the linear system merge into pure chiral states. Moreover, careful analysis of the full laser rate equations showed that the gain nonlinearity directs the effective relative loss between the lasers towards its value at the EP, making the chirality of the system remarkably resilient to loss and detuning noises. 
The presented approach is scalable and can be applied to any number of lasers, to various network geometries and other coupled-laser systems. It may lead to new insights in topology \cite{TIL}, complex band structures \cite{Reuter2016}, and spin simulators \cite{McMahonaah5178}. Moreover, since the complex potential is on-site, the concept can be transferred relatively easily to completely different physical systems.

The authors thank Oren Raz and Nitsan Bar for fruitful discussions.

\bibliography{main_bib}

\begin{thebibliography}{31}
\bibitem [{det({\natexlab{a}})}]{detunChirality}%
  \BibitemOpen
\bibfield  {journal} {  }\bibinfo {note} {{F}or coupled lasers, applying only
  frequency detuning may lead to the emergence of chiral modes in some cases.
  However, the mechanism remains similar: the oscillation amplitude of one
  laser is reduced (due to nonlinearity) and thus weakens the indirect link
  \cite{Supp}}\BibitemShut {NoStop}%
\bibitem [{det({\natexlab{b}})}]{detuning}%
  \BibitemOpen
  \bibinfo {note} {{T}he frequency detuning $\Delta\Omega$ is the phase that
  the light accumulates after impinging the SLM (in radians). It can be
  translated to units of frequency by multiplying $\Delta\Omega$ with
  $\frac{FSR}{2\pi}$, where FSR stands for Free Spectral Range of the lasers,
  which is $60$MHz}\BibitemShut {NoStop}%
\bibitem [{lin()}]{linearSteadyState}%
  \BibitemOpen
  \bibinfo {note} {{D}ue to the dissipative nature of the Hamiltonian, steady
  state solution other than $\boldsymbol{E}=0$ can be obtain only if sufficient
  gain is introduced, but for the linear analysis at this step, only the
  eigenmodes of eq. 2 are of interest}\BibitemShut {NoStop}%
\end{thebibliography}

\end{document}